\documentclass[twocolumn]{revtex4}

\usepackage{amsmath}
\usepackage{graphicx}
\usepackage{amsfonts}
\usepackage{color}

\begin{document}

\title{Quantum learning robust to noise}
\author{Andrew W. Cross}
\email{awcross@us.ibm.com}
\affiliation{IBM T. J. Watson Research Center, 1101 Kitchawan Road, Yorktown Heights, NY 10598}

\author{Graeme Smith} 
\email{gsbsmith@gmail.com}

\affiliation{IBM T. J. Watson Research Center, 1101 Kitchawan Road, Yorktown Heights, NY 10598}
\author{John A. Smolin}
\email{smolin@us.ibm.com}
\affiliation{IBM T. J. Watson Research Center, 1101 Kitchawan Road, Yorktown Heights, NY 10598}
\date{18 July 2014}

\begin{abstract}
  Noise is often regarded as anathema to quantum computation, but in
  some settings it can be an unlikely ally. We consider the problem of
  learning the class of $n$-bit parity functions by making queries
  to a quantum example oracle. In the absence of noise, quantum and
  classical parity learning are easy and almost equally powerful, both
  information-theoretically and computationally.  We show that in
  the presence of noise this story changes dramatically.  Indeed, the
  classical learning problem is believed to be intractable, while the quantum version
  remains efficient.  Depolarizing the qubits at the oracle's output at any constant
  nonzero rate does not increase the computational (or query) complexity of
  quantum learning more than logarithmically. However, the problem of learning from corresponding classical examples
  is the Learning Parity with Noise (LPN) problem, for which the best known algorithms have superpolynomial complexity.   This creates the possibility of observing a quantum advantage with a few hundred noisy qubits.  The presence of noise is essential for creating this
  quantum-classical separation.
\end{abstract}

\maketitle

\section{Introduction}

A theory of quantum fault-tolerance has been erected to
overcome pervasive decoherence
\cite{abo98,kit97,agp06}. Without such
fault-tolerant machinery, large classes of quantum algorithms can fail
to give any significant improvement over classical algorithms
\cite{regev08}. This may lead one to suppose that noise can only rob
quantum algorithms of their supremacy or at best increase the cost of
running them. Here we exhibit a problem for which noise is instead a
more significant classical foe and is crucial to achieving a quantum
speed-up, or rather, a classical slow-down.  This challenges the received wisdom that quantum computations 
are inherently delicate while classical computation is more robust.

We consider the problem of learning a class of Boolean functions by
making queries to a quantum example oracle \cite{bshouty99}. Such an
 oracle provides a quantum state that encodes a hidden function, and the goal is to discover the function efficiently, meaning
 with a number of queries and an  amount of post-processing that scales polynomially in the number of input bits. In the quantum setting, we are permitted to apply coherent operations to the quantum state, whereas in the classical setting we must first measure the state in the computational basis before further computation. This model of quantum learning differs from other attempts to use quantum computers to perform machine learning tasks \cite{anguita03,pudenz13,lloyd13,rebentrost13,wiebe14}. Information-theoretically, quantum learning from queries to ideal oracles is only polynomially more powerful than classical learning \cite{servedio04,atici05}. Computationally, however, there is a class of functions that is polynomial time learnable from quantum coherent queries but not from classical queries, under the assumption that factoring Blum integers is intractable \cite{servedio04}.

In this work, we exhibit a learning problem with a superpolynomial quantum computational speed-up only in the presence of noise.  The physical implementation of any oracle on bare qubits will inevitably be noisy.  To fairly assess the performance of a quantum algorithm given access to such an oracle, we must compare it to a classical algorithm given access to a noisy classical oracle with similar noise characteristics. To do this, we imagine constructing a noisy classical oracle by completely dephasing the inputs and outputs of a noisy quantum oracle in the computational basis. 

For the class of parity functions, we show that depolarizing
the example oracle's output at any nonzero rate has a small (logarithmic) effect on
the computational complexity of learning from quantum coherent
examples. However, the function cannot be learned from classical
examples provided by the corresponding noisy classical oracle, as this is equivalent to a problem called Learning Parity with Noise (LPN), for which the best known algorithm has superpolynomial complexity \cite{lyubash05}. Both problem settings are tractable without noise, so a quantum advantage is not merely retained; it occurs because of the noise.

The rest of the paper is organized as follows.  Section~\ref{sec:def} reviews definitions relevant to quantum
learning. Sections~\ref{sec:lp} and \ref{sec:lpn} consider the
learning problem without and with noise, respectively. Finally,
Section~\ref{sec:conc} concludes.

\section{Definitions}\label{sec:def}

We begin by reviewing relevant definitions. A \emph{membership oracle} for a Boolean function $f:\{0,1\}^n\rightarrow \{0,1\}$ is an oracle that, when
queried with input $x$, outputs the result $f(x)$.  It is so called because we can think of $f(x)$ as telling us whether input $x$ belongs to 
a set associated with the function (namely, the set of inputs that evaluate to 1).  A query to a \emph{uniform random example oracle} for a Boolean function $f$ returns an ordered pair $(x,f(x))$ where $x$ is drawn uniformly at random from the set of all possible inputs of $f$. The membership oracle gives an agent freedom to choose the input, whereas the example oracle merely allows one to ``push a button'' and request an output.

\begin{figure}[h]
\includegraphics[width=3in]{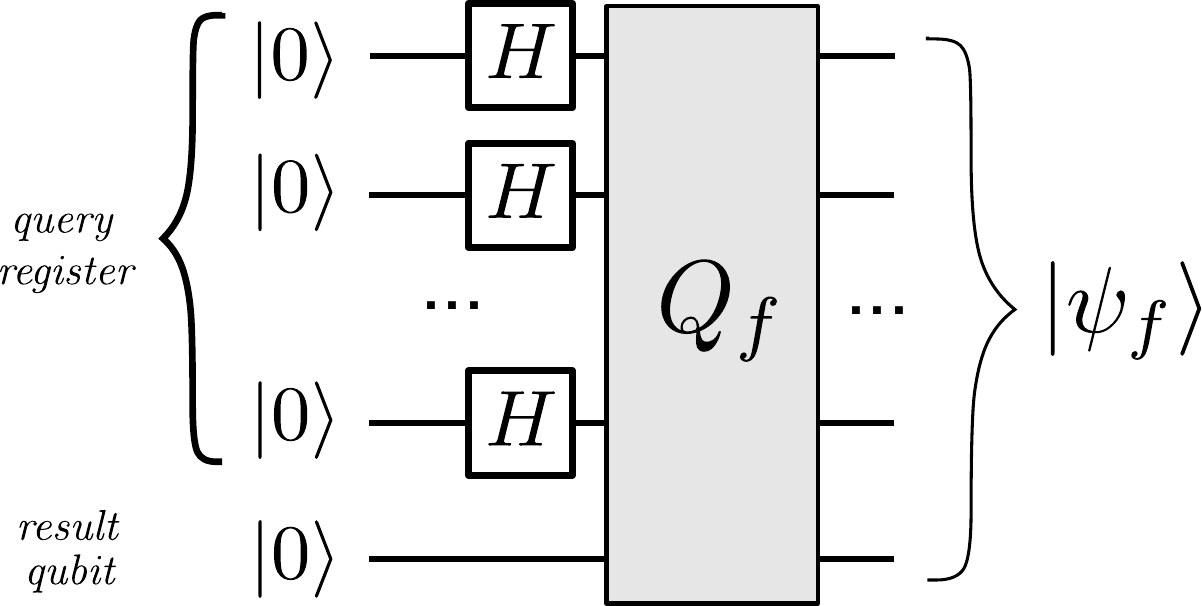}
\caption{One can construct a uniform quantum example oracle from a quantum membership oracle for $f$. $H$ denotes a Hadamard gate.  The quantum learner then performs a quantum computation to identify the function $f$. The corresponding classical example oracle is obtained from the quantum oracle by measuring its output qubits in their computational
bases.  The classical learner uses this output together with classical computation to learn the function.\label{fig:qex}}
\end{figure}

The problem of learning a class of Boolean functions by querying such oracles can be generalized to a quantum coherent setting \cite{bshouty99}.
A \emph{quantum membership oracle} $\textsc{Q}_f$ is a unitary transformation that acts on the computational basis states as
\begin{equation}
\textsc{Q}_f: |x,b\rangle\mapsto |x,b\oplus f(x)\rangle,
\end{equation}
where $x\in \{0,1\}^n$ and $b\in \{0,1\}$. A \emph{uniform quantum example oracle} for $f$ outputs the quantum state
\begin{equation}
|\psi_{f}\rangle\equiv\frac{1}{2^{n/2}}\sum_{x\in\{0,1\}^n} |x,f(x)\rangle.
\end{equation}
This oracle only gives the learner freedom to request some number of quantum states, each at unit cost. For both oracles, the \emph{query register} comprises the qubits containing $x$, and the \emph{result qubit} is the auxiliary qubit containing $f(x)$ (Fig.~\ref{fig:qex}).

Given any quantum oracle, we define a corresponding classical oracle by completely dephasing every interface to the quantum oracle, passing each input/output qubit through a channel ${\cal E}_Z(\rho)=(\rho+Z\rho Z)/2$, where 
$Z = \left( \begin{matrix}1 & 0 \\ 0 & -1 \end{matrix}\right)$. Any quantum membership oracle becomes a classical membership oracle, and any uniform quantum example oracle becomes a uniform random example oracle. This definition allows us to begin with a noisy quantum oracle and identify a corresponding noisy classical oracle with similar noise characteristics \footnote{Equivalently, one can instead completely dephase the {\em learner's} interface by moving the dephasing channels outside the quantum oracle. One can then define a classical learner as a learner who interacts with oracles through dephased interfaces, whereas a quantum learner has no such restriction. Classical and quantum learners can now be given access to the same noisy quantum oracle.}.

In the domain of learning theory, a \emph{concept} $f$ is a Boolean function $f:\{0,1\}^n\rightarrow \{0,1\}$. A \emph{concept
  class} ${\mathfrak C}=\cup_{n\geq 1} C_n$ (hereafter, \emph{class})
is a collection of concepts, and each $C_n$ contains the concepts whose domain is $\{0,1\}^n$.
Given a \emph{target concept} $f\in {\mathfrak C}$, a typical goal is
to construct a {\it hypothesis} function $h:\{0,1\}^n\rightarrow
\{0,1\}$ that agrees with $f$ on at least a $1-\epsilon$ fraction of
the inputs in $\{0,1\}^n$, i.e.
\begin{equation}
\mathrm{Pr}_{x}\left[h(x)=f(x)\right]\geq 1-\epsilon
\end{equation}
where $x$ is drawn from the uniform distribution.
Such a function is called an {\it $\epsilon$-approximation of $f$}.

A class ${\mathfrak C}$ is \emph{efficiently PAC (Probably Approximately Correct \cite{valiant84}) learnable under the uniform distribution} if given a uniform example oracle for any target concept $f\in {\mathfrak C}$, there is an algorithm that
\begin{enumerate}
\item for any $\epsilon,\delta\in (0,1/2)$, outputs an $\epsilon$-approximation $h$ of $f$ with probability $1-\delta$,
\item runs in time and uses a number of queries that is $\mathrm{poly}(n,1/\epsilon,1/\delta)$.
\end{enumerate}
The definition of learning is identical in the quantum setting except that the uniform example oracle is replaced by a uniform quantum example oracle, and the allowed computations may be coherent. We will also consider example oracles corrupted by noise of constant rate $\eta<1/2$ in a way that is defined later. The definition of learning is unchanged in this case, although one may require the algorithm to run in time $\mathrm{poly}(1/(1/2-\eta))$ as well.

We now restrict the discussion to the class of parity functions
\begin{equation}
f_a(x) = \langle a,x\rangle = \sum_{j=1}^n a_jx_j\ \mathrm{mod}\ 2
\end{equation}
where $a\in \{0,1\}^n$ and $a_j$ ($x_j$) denotes the $j$th bit of $a$
($x$). We are given access to a uniform quantum example oracle for the
unknown concept $f_a(x)=\langle a,x\rangle$. 
If we incorrectly guess even a
single bit of $a$, our hypothesis function is a $1/2$-approximation to
$f_a$ and remains so for any number of incorrect bits. Therefore, we
must with high probability find $a$ exactly, and this is what we require hereafter.

\section{Learning from ideal queries}\label{sec:lp}

First, consider the noiseless case where each query returns a pure
quantum state. This case is tractable for both quantum and classical
queries as we now review.

The classical oracle provides an example $(x,f_a(x))$ where $x$ is uniformly random over $\{0,1\}^n$. Since
$f_a(x)$ is a linear function, it is clear that $n$ queries are
sufficient to learn $f_a$ exactly with a constant probability of
success. The probability that $n$ queries produce linearly independent
examples is
\begin{equation}
\prod_{j=0}^{n-1}(1-2^{j-n}),
\end{equation}
which is greater than $1/4$ for any $n>1$. Any algorithm that detectably fails with constant probability less than $p$ and otherwise succeeds can be repeated no more than $\log_{1/p}(1/\delta)$ times to reduce the failure probability below $\delta$. The value of $a$ is
obtained from the examples by Gaussian elimination.

In the quantum setting, $f_a$ can be learned with constant probability
from a single query. Given $|\psi_f\rangle$, apply Hadamard gates
\begin{equation}
H=\frac{1}{\sqrt{2}}\left(\begin{array}{cc}
1 & 1 \\
1 & -1 \end{array}\right)
\end{equation}
to each of the $n+1$ output qubits. A simple calculation shows that the output state becomes
\begin{equation}
\frac{1}{\sqrt{2}}\left(|0^n,0\rangle+|a,1\rangle\right).
\label{eq:outputstate}
\end{equation}
Therefore, with probability $1/2$, measurement reveals the value of
$a$ directly in the query register whenever the result qubit is one. Again the probability of success can be amplified with $O(\log(1/\delta))$ queries.

Note that this is very similar to the Bernstein-Vazirani
algorithm \cite{BV} but adapted to use an example oracle rather than a membership oracle.  The only difference is our treatment of the final qubit.   In the Bernstein-Vazirani algorithm, the result qubit is input as a $|-\rangle$ state and will, with certainty in the noiseless case, end up as a $|1\rangle$.  It then does not even need to be measured.  For the example oracle we have considered, we do not have the luxury of choosing the input, so we simply check that the result qubit is $|1\rangle$, which collapses the output state of the other $n$ qubits to the result of the Bernstein-Vazirani oracle.

\section{Learning in the presence of noise}\label{sec:lpn}

Now we consider how the situation changes when we add noise to the
output of the example oracle.  We will see that learning parity from a noisy example oracle 
seems to become computationally intractable, while the same task with a noisy quantum 
example oracle can be solved efficiently on a quantum computer.  We will first consider a simple
case that is easy to analyze followed by the more realistic case of
depolarizing noise.

\subsection{Classification noise}

Just about the most trivial model of noise one can imagine is this
one: flip the result qubit with probability $\eta$ by applying
the Pauli $\sigma_x={0\ 1 \choose 1\ 0}$ operator. Classically, learning $f_a$ from such corrupted results is called \emph{learning parity with noise} (LPN). 
This problem is an average-case version \cite{lyubash05} of the NP-hard problem of decoding a linear code \cite{berlekamp78}, which is also known to be hard to
approximate \cite{hastad97}. The LPN problem
is believed to be computationally intractable. Potential cryptographic
applications have been proposed for this problem and its
generalizations \cite{regev05}, and the best known algorithms for LPN
are sub-exponential (but
super-polynomial) \cite{blum03,lyubash05}. Problem instances with
hundreds of bits may be impractical to solve \cite{lev06}.

However, the quantum case remains easy. With noise, the output of the oracle transformed by Hadamards becomes the mixture of
(\ref{eq:outputstate}) with probability $1-\eta$ and
\begin{equation}
\frac{1}{\sqrt{2}}(|0^n,1\rangle+|a,0\rangle) 
\end{equation}
with probability $\eta$. The probability that the query register contains $a$ remains $1/2$,
independent of $\eta$.  Thus, after $k$ queries, the probability of observing $a$ is $1-(1/2)^k$.  This suggests the simple strategy
of reporting either $a=$``whatever nonzero result is seen'' or $0^n$ otherwise. It fails with a probability that is exponentially small in $k$ and independent of $n$.

This strategy is strictly suboptimal since it ignores the information contained in the result qubit.  The fact that it works so well, regardless, suggests that our noise model is rather unfair to the classical case by degrading the result qubit, which the quantum
algorithm hardly even needs.  Indeed, in the Bernstein-Vazirani algorithm, the output bit isn't measured at
all. Still, this simple case serves to illustrate how noise can more severely impact the classical learner. Even in the more realistic noise model we consider next, the quantum algorithm survives the addition of significant amounts of noise because the quantum queries reveal so very much.

\subsection{Depolarizing noise}

Now we consider the case where the output of the oracle is subject to
independent depolarizing noise $D_\eta^{\otimes (n+1)}$ where
$D_\eta(\rho) = (1-2\eta)\rho + 2\eta I/2$ and $\eta<1/2$ is a
constant known noise rate. This noise process is an idealization of
realistic independent noise and corrupts the ideal output of the oracle with probability proportional to $\eta$.

Classically, in the presence of this noise, we obtain examples $(x\oplus
e_{1:n},f_a(x)\oplus e_{n+1})$ where $x$ is uniformly random over
$\{0,1\}^n$ and each bit of the noise $(e_{1:n},e_{n+1})$ is $1$
with probability $\eta$. Here $e_{1:n}\in \{0,1\}^n$ and
$e_{n+1}\in\{0,1\}$. Since $x$ is uniformly random, $x'=x\oplus
e_{1:n}$ is uniformly random as well and
\begin{equation}
(x\oplus e_{1:n},f_a(x)\oplus e_{n+1})=(x',f_a(x'\oplus e_{1:n})\oplus e_{n+1}).
\end{equation}
The probability that $f_a(x'\oplus e_{1:n})\neq f_a(x')$ depends on the value of $a$ and is given by
\begin{equation}
\zeta_a=\sum_{w=1}^n\sum_{k=1,\textrm{odd}}^w {|a|\choose k}{n-|a|\choose w-k}\eta^w(1-\eta)^{n-w}
\end{equation}
where $|a|$ denotes the number of $1$'s in $a$ (also called the Hamming weight). The total probability
of error on the result bit is simply
$\eta':=\eta(1-\zeta_a)+\zeta_a(1-\eta)$. Therefore, learning from these examples is the LPN problem with noise rate $\eta'\in [\eta,1-\eta]$.

In contrast, coherent manipulation of the noisy output state of the
example oracle allows a quantum learner to learn $f_a$ in a number of queries that is logarithmic in $n$. The algorithm is as follows. Make $k=O(\log n)$ queries to the example oracle, and for each query, Hadamard transform all $n+1$ noisy output qubits and measure them to obtain an outcome. Each outcome has the form $(m,b)$ where $m\in \{0,1\}^n$ is a result string in the query register and the result bit $b$ is uniformly random. Discard the outcome if $b=0$ and otherwise retain the result string $m$. We are left with $k'$ result strings $m_1$, $m_2$, $\dots$, $m_{k'}$ on which we perform a bit-wise majority vote  to obtain an estimate $\hat{a}$ of $a$.

We will now argue that the estimate $\hat{a}$ obtained from this protocol is equal to $a$ for appropriately chosen parameters.
Given any constant $\delta>0$, the algorithm must find an estimate $\hat{a}$ such that $\mathrm{Pr}[\hat{a}\neq a]<\delta$.
We make repeated use of a loose form of the Chernoff bound
\begin{equation}
\mathrm{Pr}\left[|X-\eta k|<\delta\eta k\right]
> 1-2e^{-\delta^2\eta k/3}\equiv 1 - B_k(\eta,\delta)
\label{eq:Chernoff}
\end{equation}
where $X$ is the sum of $k$ independent Bernoulli random variables with $\mathrm{Pr}(1)=\eta$ and $0<\delta<1$. Clearly we can query the oracle until we retain a total of $k'$ result strings.  This takes $2k'$ expected queries.  By performing, say, $3k'$ queries, 
we are guaranteed via Eq.~(\ref{eq:Chernoff}) to retain fewer than $k'$ with probability exponentially small in $k'$.  Now, let $D_q^\eta$ be the probability distribution over $\{0,1\}^n$ that corresponds to the bit string $q$ corrupted by independent bit-flip noise of rate $\eta$. The retained strings are drawn from the distribution $(1-\eta)D_a^\eta+\eta D_{0^n}^\eta$. Let $s$ be the random variable giving the unknown number of strings drawn from $D_a^\eta$. These successful queries, which 
we take to be $m_1,m_2,\dots,m_s$ without loss of generality,  contain information about the hidden function.  The expected value of $s$ is $\mu_s=(1-\eta)k'$, and its variation from this mean is controlled by
\begin{align}
\mathrm{Pr}\left[|s-\mu_s|<\delta'\mu_s\right]
> 1 - B_{k'}(1-\eta,\delta').
\end{align}
Our algorithm votes independently on the $j$th bits of the strings for each $j=1,2,\dots, n$. Let $M_j$ be the random variable corresponding to the sum
$(m_1)_j+(m_2)_j+\dots+(m_{k'})_j$
of the $j$th bits. The worst case occurs when $a_j=1$ and we assume this. Define random variables $M_j^{(a)}=m_1+\dots+m_s$ and $M_j^{(0)}=M_j-M_j^{(a)}$ with means $\mu_j^{(a)}=(1-\eta)s$ and $\mu_j^{(0)}=\eta(k'-s)$, respectively. The mean of $M_j$ is $\mu_j=\mu_j^{(a)}+\mu_j^{(0)}$. Conditioned on obtaining a typical value of $s$, the probability of a successful vote on the $j$th bit is
\begin{align}
\gamma_j\geq & \mathrm{Pr}\left[M_j>\frac{k'}{2}\right] 
 \geq \mathrm{Pr}\left[|M_j-\mu_j|<\delta'\mu_j\right]\label{eq:line} \\
\geq & \mathrm{Pr}\left[|M_j^{(a)}-\mu_j^{(a)}|<\delta'\mu_j^{(a)}\right]\times \\
  & \mathrm{Pr}\left[|M_j^{(0)}-\mu_j^{(0)}|<\delta'\mu_j^{(0)}\right] \\
> & 1 - 2B_{k'}(\tilde{\eta},\delta').\label{eq:line2}
\end{align}
We have defined $\tilde{\eta}=\eta(1-(1+\delta')(1-\eta))$ and chosen $\delta'<\eta/(1-\eta)$.  The second inequality of $(\ref{eq:line})$ follows by further choosing $1-\delta'>\frac{1}{2}((1-2\eta)(1-\eta)+\eta)^{-1}$. To find $(\ref{eq:line2})$, we used $\tilde{\eta}\leq(1-\delta')(1-\eta)^2$.  This gives us an upper bound
\begin{align}
\mathrm{Pr}\left[\hat{a}_j \neq a_j\right] = 1-\gamma_j < 2 B_{k'}(\tilde{\eta},\delta'),
\end{align}
on the probability of the $j$th bit being computed incorrectly, which we can use together with a union bound to find
\begin{align}
\mathrm{Pr}\left[\hat{a} \neq a\right]  \leq \sum_j \mathrm{Pr}\left[\hat{a}_j \neq a_j\right] 
 < 2n   B_{k'}(\tilde{\eta},\delta').
\end{align}
Choosing 
\begin{align}
k' > \frac{3}{(\delta')^2\tilde{\eta}}\log\left(\frac{4n}{\delta}\right)
\end{align}
ensures that $\mathrm{Pr}\left[\hat{a} \neq a\right] < \delta$. For sufficiently large $\eta$, one can verify easily that $(\delta')^2\tilde{\eta}$ is a polynomial in $(1/2-\eta)$, and therefore $k'=O(\mathrm{poly}(1/(1/2-\eta)))$.

\section{Conclusion}\label{sec:conc}

We have defined the problem of quantum learning from a noisy quantum
example oracle and shown that the class of parity functions can
be learned in logarithmic time from corrupted quantum
queries.  In contrast, it appears to be intractable to learn this class in polynomial time from classical queries to the corresponding classical noisy oracle \footnote{In fact, even a quantum computer given access to this classical noisy example 
oracle is not known to be able to efficiently learn this class.}.  
If the oracle is ideal, the problem is tractable for both quantum and
classical learners, so the noise plays an essential role in the
exhibited behavior. For this problem at least, decoherence is an ally
of quantum computation.

The example oracle for parity can be implemented in practice with
$O(n)$ one- and two-qubit gates.  The quantum learner then needs
only single-qubit gates and measurements, or even just measurements in
a nonstandard basis.  This suggests that an experimental demonstration
may be quite practicable. The independent depolarizing noise model we
use is an idealization of realistic decoherence.  Although a more
detailed study of actual experimental noise would be needed, it
should be possible to demonstrate quantum supremacy for learning using several hundred \emph{noisy} qubits, i.e. without the use of quantum error-correction.  In the meantime, since a classical learner requires at least $n$ queries in the noiseless case while the quantum learner needs only $O(\log n)$ queries with
noise, a quantum advantage for query-complexity, while small, could
be shown experimentally in existing systems.

\begin{acknowledgements}
We thank Charlie Bennett, Sergey Bravyi, and Martin Suchara for helpful comments. AWC and JAS acknowledge support from IARPA under contract W911NF-10-1-0324. JAS and GS acknowledge NSF grant CCF-1110941.
\end{acknowledgements}

\bibliographystyle{unsrt}

\begin{thebibliography}{10}

\bibitem{abo98}
D.~Aharonov and M.~Ben-Or.
\newblock Fault tolerant quantum computation with constant error.
\newblock {\em Proceedings, 29th Annual ACM Symposium on Theory of Computing},
  pages 176--188, 1998.

\bibitem{kit97}
A.~Yu. Kitaev.
\newblock Quantum computations: algorithms and error correction.
\newblock {\em Russian Math. Surveys}, 52:1191--1249, 1997.

\bibitem{agp06}
P.~Aliferis, D.~Gottesman, and J.~Preskill.
\newblock Quantum accuracy threshold for concatenated distance-3 codes.
\newblock {\em QIC}, 6(97), 2006.

\bibitem{regev08}
O.~Regev and L.~Schiff.
\newblock Impossibility of a quantum speed-up with a faulty oracle.
\newblock {\em Proc. 35th International Colloquium on Automata, Languages, and
  Programming (ICALP)}, pages 773--781, 2008.

\bibitem{bshouty99}
N.~Bshouty and J.~C. Jackson.
\newblock Learning dnf over the uniform distribution using a quantum example
  oracle.
\newblock {\em SIAM J. Comput.}, 28(3):1136--1153, 1999.

\bibitem{anguita03}
D.~Anguita, S.~Ridella, F.~Rivieccio, and R.~Zunino.
\newblock Quantum optimization for training support vector machines.
\newblock {\em Neural Networks}, 16(5--6):763--770, 2003.

\bibitem{pudenz13}
K.~Pudenz and D.~Lidar.
\newblock Quantum adiabatic machine learning.
\newblock {\em Quant. Inf. Proc.}, 12(5):2027--2070, 2013.

\bibitem{lloyd13}
S.~Lloyd, M.~Mohseni, and P.Rebentrost.
\newblock Quantum algorithms for supervised and unsupervised machine learning.
\newblock {\em arXiv:1307.0411}, 2013.

\bibitem{rebentrost13}
P.~Rebentrost, M.~Mohseni, and S.~Lloyd.
\newblock Quantum support vector machine for big data classification.
\newblock {\em arXiv:1307.0471}, 2013.

\bibitem{wiebe14}
N.~Wiebe, A.~Kapoorb, and K.~Svore.
\newblock Quantum nearest-neighbor algorithms for machine learning.
\newblock {\em arXiv:1401.2142}, 2014.

\bibitem{servedio04}
R.~Servedio and S.~Gortler.
\newblock Equivalences and separations between quantum and classical
  learnability.
\newblock {\em SIAM J. Comput.}, 33(5):1067--1092, 2004.

\bibitem{atici05}
A.~Atici and R.~Servedio.
\newblock Improved bounds on quantum learning algorithms.
\newblock {\em Quantum Information Processing}, 4:355--386, 2005.

\bibitem{lyubash05}
V.~Lyubashevsky.
\newblock The parity problem in the presence of noise, decoding random linear
  codes, and the subset sum problem.
\newblock {\em Lecture Notes in Computer Science (8th International Workshop on
  Approximation Algorithms for Combinatorial Optimization Problems (APPROX))},
  3624:378--389, 2005.

\bibitem{valiant84}
L.~G. Valiant.
\newblock A theory of the learnable.
\newblock 27(11):1134--1142, 1984.

\bibitem{BV}
E.~Bernstein and U.~Vazirani.
\newblock Quantum complexity theory.
\newblock In {\em Proceedings of the Twenty-fifth Annual ACM Symposium on
  Theory of Computing}, STOC '93, pages 11--20, New York, NY, USA, 1993. ACM.

\bibitem{berlekamp78}
E.~R. Berlekamp, R.~J. McEliece, and H.~Van Tilborg.
\newblock On the inherent intractability of certain coding problems.
\newblock {\em IEEE Trans. Inf. Theo.}, 24:384--386, 1978.

\bibitem{hastad97}
J.~Hastad.
\newblock Some optimal inapproximability results.
\newblock {\em STOC}, pages 1--10, 1997.

\bibitem{regev05}
O.~Regev.
\newblock On lattices, learning with errors, random linear codes, and
  cryptography.
\newblock {\em STOC}, pages 84--93, 2005.

\bibitem{blum03}
A.~Blum, A.~Kalai, and H.~Wasserman.
\newblock Noise-tolerant learning, the parity problem, and the statistical
  query model.
\newblock {\em Journal of the ACM}, 50(4):506--519, 2003.

\bibitem{lev06}
E.~Levieil and P.~Fouque.
\newblock An improved lpn algorithm.
\newblock {\em Proc. 5th international conference on Security and Cryptography
  for Networks}, pages 348--359, 2006.

\end{thebibliography}

\end{document}